\documentclass[preprint,12pt]{elsarticle}

 \usepackage{graphics}
\usepackage{amssymb}
\biboptions{sort&compress}

\begin{document}

\begin{frontmatter}



\title{Quantitative connection between the nanoscale electronic inhomogeneity and the pseudogap of Bi$_2$Sr$_2$CaCu$_2$O$_{8+\delta}$ superconductors}


\author{Tatsuya Honma}

\address{Department of Physics, Asahikawa Medical University, Asahikawa, Hokkaido 078-8510, Japan}
\ead{honma@asahikawa-med.ac.jp}

\author{Pei Herng Hor}

\address{Department of Physics and Texas Center for Superconductivity, University of Houston,\\ Houston, Texas 77204-5005, U.S.A.}
\ead{phor@uh.edu}

\begin{abstract}
We have found a quantitative connection between the evolution of the inhomogeneous nanoscale electronic gaps (INSEG) state detected in Bi$_2$Sr$_2$CaCu$_2$O$_{8+\delta}$ by scanning tunneling microscopy/spectroscopy (STM/S) and the two universal, the upper and the lower, pseudogaps in high-temperature cuprate superconductors (HTCS). When the doping and temperature dependent INSEG map were analyzed by using our proposed hole-scale, we find that the two pseudogaps are connected to two specific coverages of the CuO$_2$ plane by INSEG: the 50\% and 100\% coverages of the CuO$_2$ planes by INSEG correspond to the upper and lower pseudogaps, respectively. This quantitative connection to the two pseudogaps indicates that the origin of the measured pseudogap energies and temperatures are intimately related to the geometrical coverage of the CuO$_2$ planes by the INSEG state. We find that INSEG and superconductivity coexist in the underdoped to the overdoped regimes. We suggest that pseudogap states are microscopically inhomogeneous and 100\% coverage of the CuO$_2$ planes by the INSEG is a necessary condition for the high-$T_c$ superconductivity.
\end{abstract}


\end{frontmatter}

% \linenumbers

\section{Introduction}
\label{}

One of the long-standing puzzles of the hole-doped high-temperature cuprate superconductors (HTCS) is the existence of the ubiquitous pseudogap state that precedes the superconducting state \cite{ito93,din96,tim99,dam03,fis07}. The pseudogap state, a partial suppression of the spectral density, generally are detected as either a pseudogap temperature ($T^*$) or a pseudogap energy ($E^*$). The initial reported $T^*$ or $E^*$ differed in details from material to material at, presumably, the same doping level and, sometimes, it is not even consistent with each other in the same material at the same doping level if determined by different experimental probes. We also showed that, using our proposed universal $P_{pl}$-scale \cite{hon04} of the planar doped-hole concentration $P_{pl}$, all measured $T^*$'s and $E^*$'s of hole-doped HTCS fell on either of the two, the upper or the lower, pseudogap lines that are independent of the number of CuO$_2$ layers in the cuprate material systems \cite{hon04}. Therefore the two universal pseudogaps are purely two-dimensional (2D)  properties of HTCS \cite{hon04,hon06,add01}. Furthermore, a unified electronic phase diagram (UEPD) was constructed in which there are four characteristic temperatures (energies) for hole-doped HTCS with an optimal superconducting transition temperature $T_c^{max}$ of $\sim$90 K, such as Bi$_2$Sr$_2$CaCu$_2$O$_{8+\delta}$, YBa$_2$Cu$_3$O$_{6+\delta}$, and HgBa$_2$CuO$_{4+\delta}$ \cite{hon08}.

\begin{figure}[b]
\includegraphics [scale=0.4]{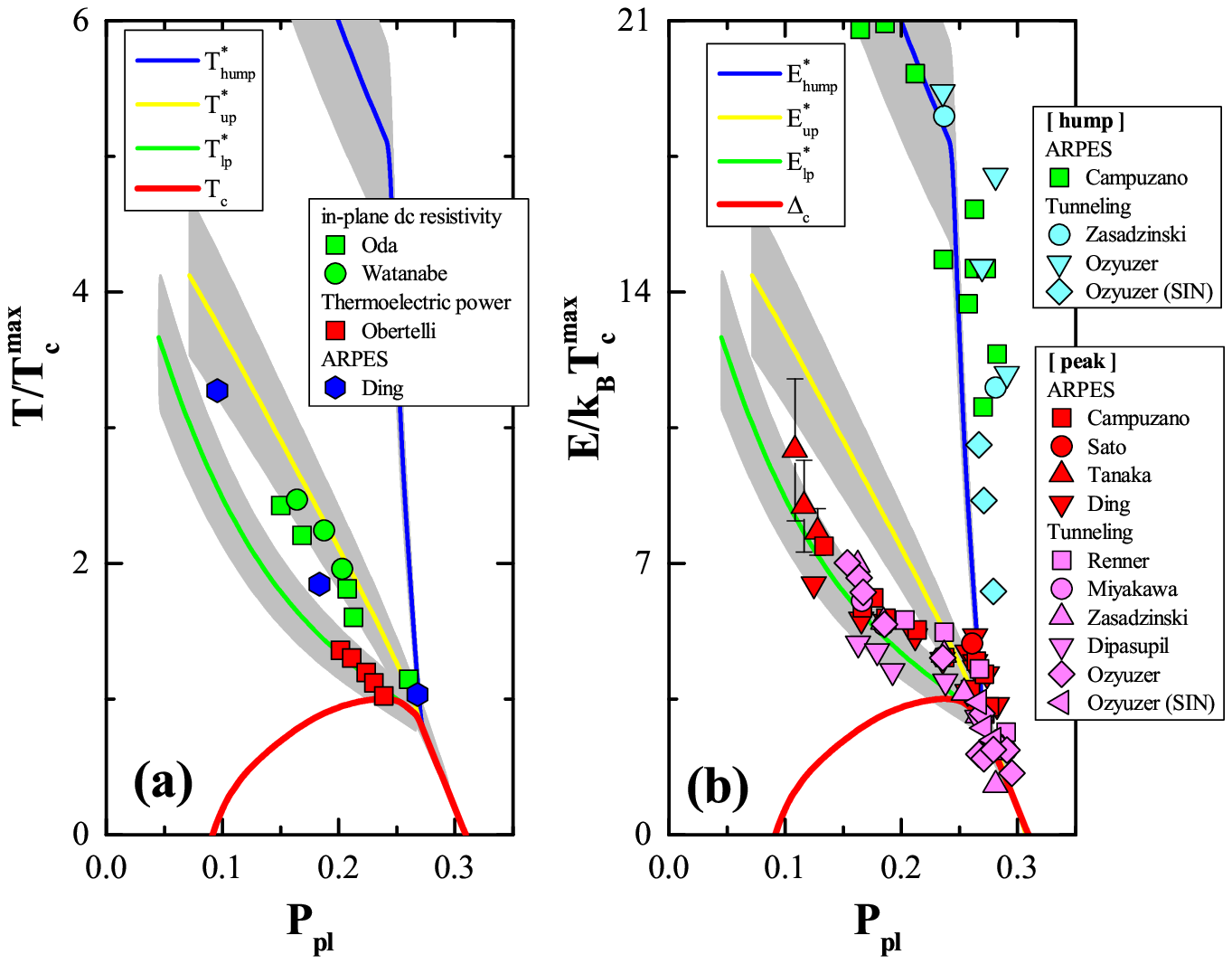}
\caption{\label{fig1} Electronic phase diagram of (a) $T/T_c^{max}-P_{pl}$ and (b) $E/k_BT_c^{max}-P_{pl}$ in Bi$_2$Sr$_2$CaCu$_2$O$_{8+\delta}$. The colored solid lines are characteristic (a) temperatures and (b) energies in the unified electronic phase diagram (UEPD) \cite{hon08}. The discrete symbols are some representative data points of Bi$_2$Sr$_2$CaCu$_2$O$_{8+\delta}$ measured by various experimental probes used to construct the UEPD. For details see text and Ref.\ \cite{hon08}, and references therein. }
\end{figure}

In Fig.\ \ref{fig1}, we plot the pseudogaps determined by various measurements of purely oxygen-doped Bi$_2$Sr$_2$CaCu$_2$O$_{8+\delta}$ together with a schematic sketch of the two universal pseudogaps and hump energy identified in the UEPD \cite{add02}. Three very distinct features can be clearly seen in Fig.\ \ref{fig1}: (i) there are three characteristic temperatures (energies): the lower pseudogap $T_{lp}^*$ ($E_{lp}^*$), the upper pseudogap $T_{up}^*$ ($E_{up}^*$) and the hump $T_{hump}^*$ ($E_{hump}^*$) in the underdoped to the slightly overdoped regimes; (ii) all three characteristic temperatures (energies) merge together with superconducting transition temperature $T_c$ (superconducting gap energy $\Delta_c$), at the slightly overdoped level; (iii) $T^*$ and $E^*$ are connected by 2$E^*/k_BT^*$ =  7 $\pm$ 1 \cite{hon08}, where $k_B$ is the Boltzmann's constant. Therefore all three characteristic temperature-scales and the three characteristic energy-scales measured by vast different experimental probes are connected by the relationship iii) and share the common electronic phase diagram, namely, UEPD.

Scanning tunneling microscopy/spectroscopy (STM/S) had made a unique contribution to the study of HTCS through the direct observation of the inhomogeneous nanoscale electronic gaps (INSEG) in the superconducting states \cite{pan01}. A recent STM/S study pushed this electronically heterogeneous picture well into the pseudogap state by showing that nanoscale local gaps persist at a temperature well above $T_c$ \cite{gom07}. It is shown that in the optimal to overdoped regimes the local INSEG appear (vanish) at a temperature $T_p$ upon cooling (warming) where $T_p$ and the INSEG energy ($\Delta_g$) are universally connected by 2$\Delta_g/k_B T_p$ =  7.9 $\pm$ 0.5 \cite{gom07}. In the underdoped regime, the situation is more complicated: two gap-like structures, the pseudogap and the pairing gap \cite{gom07,boy07}, are observed and, therefore the simple relation between local-gap vanishing temperature and gap size, namely, 2$\Delta_g/k_B T_p$ =  7.9 $\pm$ 0.5 could not be clearly pinned down. Temperature dependent STM measurements showed that the CuO$_2$ plane is gradually covered by the INSEG with decreasing temperature \cite{gom07}. Although INSEG state are considered to be related to the pseudogap and subjected to intense studies in the past decade how the temperature and doping dependence of INSEG is related to the pseudogap states were largely unexplored.

In this report we compare the temperature and the doping dependence of INSEG of purely oxygen-doped Bi$_2$Sr$_2$CaCu$_2$O$_{8+\delta}$ against UEPD \cite{hon08}. We show that in the purely oxygen-doped Bi$_2$Sr$_2$CaCu$_2$O$_{8+\delta}$ that the upper and lower pseudogaps are quantitatively connected to the specific coverage of the CuO$_2$ planes by INSEG. By comparing with the UEPD reported in Ref.\ \cite{hon08} we argue that the specific coverage of the CuO$_2$ plane by INSEG, observed in Bi$_2$Sr$_2$CaCu$_2$O$_{8+\delta}$, is a universal feature for HTCS.

\section{Analysis}

In analyzing STM data of Bi$_2$Sr$_2$CaCu$_2$O$_{8+\delta}$, we selected the single crystal data with $T_c$ reported in the publications. In general we use two criteria to extract $P_{pl}$: as the first and the most reliable method, $P_{pl}$ is determined from the value of thermoelectric power at 290 K ($S^{290}$) by using $P_{pl}$-scale \cite{hon04,hon08}. As the second method, $P_{pl}$ is determined from the value of $T_c$ by comparing it with a universal asymmetrical half-dome-shaped $T_c$-curve shown in Fig. 5 in Ref.\ \cite{hon08}. We always selected the paper that reported the value of $S^{290}$ and used the data with the value of $T_c$ when $S^{290}$ is not available. In this STM analysis, we could not find the data with $S^{290}$. So, we adopted the second criteria. For some YBa$_2$Cu$_3$O$_{6+\delta}$, the value of $P_{pl}$ was estimated from the double plateau $T_c$-curve of Fig. 2(a) of Ref. \ \cite{hon07}. For HgBa$_2$CuO$_{4+\delta}$, the value of $P_{pl}$ was estimated from a relation of $T_c$ versus $P_{pl}$ based on $T_c$ versus $S^{290}$ extracted from Ref.\ \cite{yam00}.

\section{Results and discussion}

\begin{figure}[b]
\includegraphics [scale=0.4]{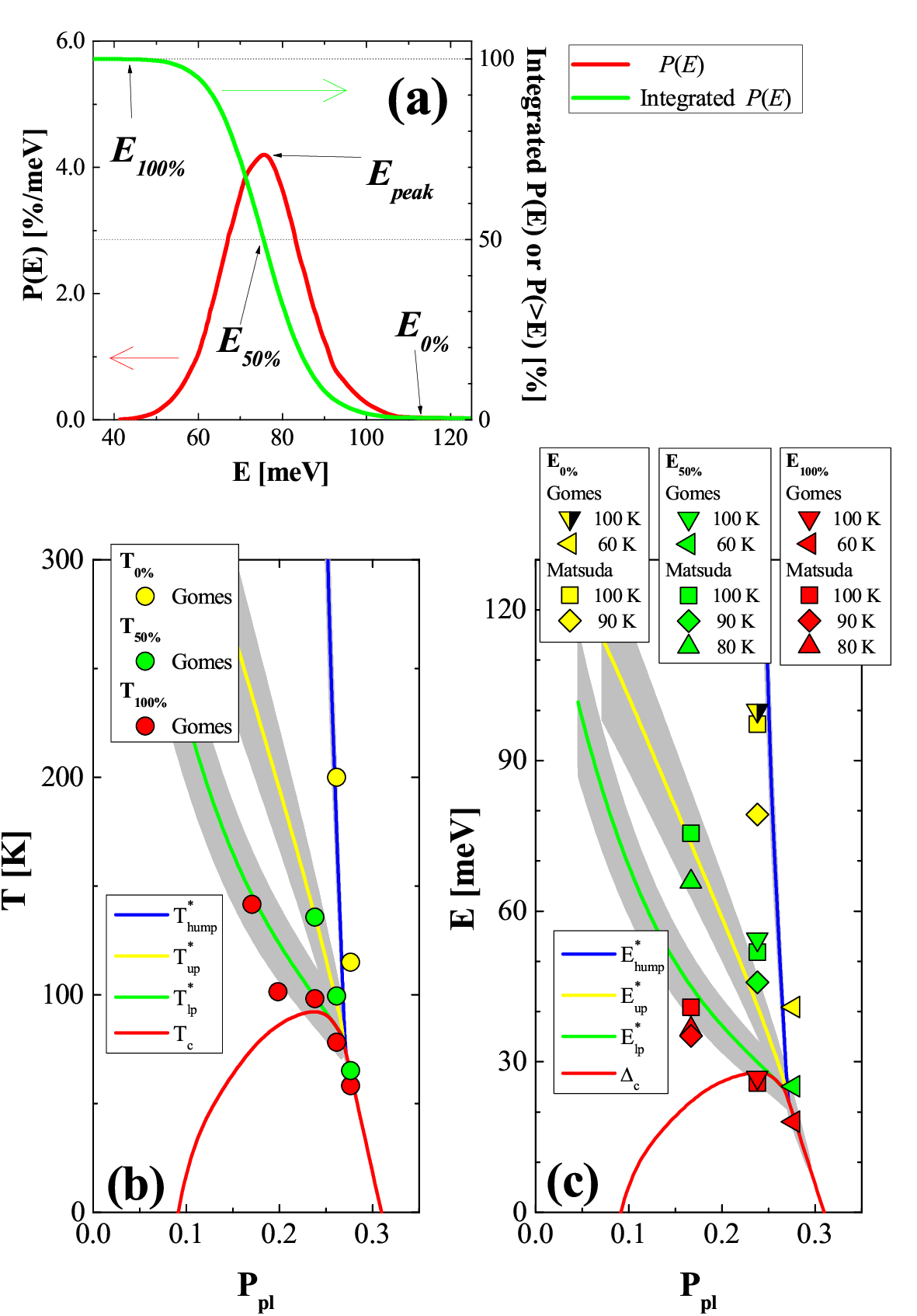}
\caption{\label{fig2} (a) Operational definition of $E_{0\%}$, $E_{50\%}$ and $E_{100\%}$. For details see text. (b) $T_{0\%}$, $T_{50\%}$ and $T_{100\%}$ versus $P_{pl}$ of Bi$_2$Sr$_2$CaCu$_2$O$_{8+\delta}$. The plotted data are coming from Ref.\ \cite{gom07}. (c) $E_{0\%}$, $E_{50\%}$ and $E_{100\%}$ versus $P_{pl}$ of Bi$_2$Sr$_2$CaCu$_2$O$_{8+\delta}$. The data are coming from Refs.\ \cite{gom07,mat03}. $E_{0\%}$ at 100 K by Gomes is a lower bound of $E_{0\%}$. }
\end{figure}

To compare the temperature and doping evolution of INSEG, upon cooling or warming, with the pseudogaps in Bi$_2$Sr$_2$CaCu$_2$O$_{8+\delta}$ we define the temperatures corresponding to 0\%, 50\% and 100\% coverage of the CuO$_2$ planes by INSEG as $T_{0\%}$, $T_{50\%}$ and $T_{100\%}$, respectively. Similarly for CuO$_2$ planes that are completely covered by the INSEG we define the energies corresponding to 0\%, 50\% and 100\% coverage of the CuO$_2$ planes by INSEG as $E_{0\%}$, $E_{50\%}$ and $E_{100\%}$, respectively. Both $T_{0\%}$ and $T_{100\%}$ are determined from the experimentally observed distribution curve by Gomes $et$ $al$. \cite{gom07}.  For all STM data we analyzed, we confirmed that INSEG distributions are Gaussian suggesting that the gap distribution is driven by randomness. The value of squared multiple correlation coefficient adjusted for the degrees of freedom was always over 0.97, except of some Gaussian fittings of 0.9 $\sim$ 0.95.

In Fig.\ \ref{fig2}(a), we plot the red curve as the probability $P$($E$) of finding a nanoscale gap at energy $E$ and the green curve as the probability $P$($>E$) to find a nanoscale gap that is larger than $E$ of a typical INSEG map. The percentage of the area covered by the gaps larger than $E$ out of the total gapped area is calculated by integrating $P$($E$) from $E$ [meV] to $\infty$  [meV] (0 $\leq E < \infty$). The intersection of green curve with $P$($>E$) = 0\%, 50\% and 100\% are corresponding to $E_{0\%}$, $E_{50\%}$ and $E_{100\%}$, respectively. While theoretically $E_{0\%}$ and $E_{100\%}$ should correspond to $E$ = $\infty$ and 0, respectively both $E_{0\%}$ and $E_{100\%}$ we plotted are read directly from the gap distribution observed in Bi$_2$Sr$_2$CaCu$_2$O$_{8+\delta}$ by two groups \cite{gom07,mat03}. Therefore they do not correspond to the theoretical  end points of the Gaussian distribution.

In Fig.\ \ref{fig2}(b), we plot the temperatures $T_{0\%}$, $T_{50\%}$ and $T_{100\%}$, as a function of $P_{pl}$, determined from the temperature dependence of gap distribution reported by Gomes $et$ $al$. \cite{gom07}. It is clearly seen that $T_{100\%}$ and $T_{50\%}$ correspond to $T_{lp}^*$ and $T_{up}^*$, respectively. At the slightly overdoped regime, $T_{hump}^*$ is associated with the onset temperature, the $T_{0\%}$, of the INSEG. In Fig.\ \ref{fig2}(c) with an error band defined by 2$E^*/k_BT^*$ = 7 $\pm$ 1, we plot the energies, $E_{0\%}$, $E_{50\%}$ and $E_{100\%}$, extracted from the gap distribution measured at 100 K, 90 K, 80 K and 60 K by two groups \cite{gom07,mat03}. Note that in order to extract the corresponding energies for 0\%, 50\% and 100\% coverage of the CuO$_2$ planes, we have to use the gap map that has completely covered the CuO$_2$ planes since in order to determine 50\% coverage, we first need to know the 100\% coverage. Therefore, all the subsequent gap distribution data we analyzed are collected below $T_{100\%}$ = $T_{lp}^*$.  It is clearly seen that, from the underdoped regime to the slightly overdoped regime, $E_{100\%}$ and $E_{50\%}$ correspond to $E_{lp}^*$ and $E_{up}^*$, respectively. At the slightly overdoped regime, $E_{0\%}$ lies on $E_{hump}^*$. Here is one of the most important conclusions of this paper, namely, in Bi$_2$Sr$_2$CaCu$_2$O$_{8+\delta}$, we found that the temperature and doping dependent coverage of CuO$_2$ planes by INSEG is intrinsically connected to the electronic phase diagram shown in Fig.\ \ref{fig1} where 50\% and 100\% coverages correspond to upper and lower pseudogaps, respectively. Since the upper and lower pseudogaps are pure 2D properties \cite{hon04}, the quantitative connection between the INSEG state coverage to both pseudogap states over wide doping ranges strongly suggests that the INSEG state is also a 2D property.

\begin{figure}[b]
\includegraphics [scale=0.4]{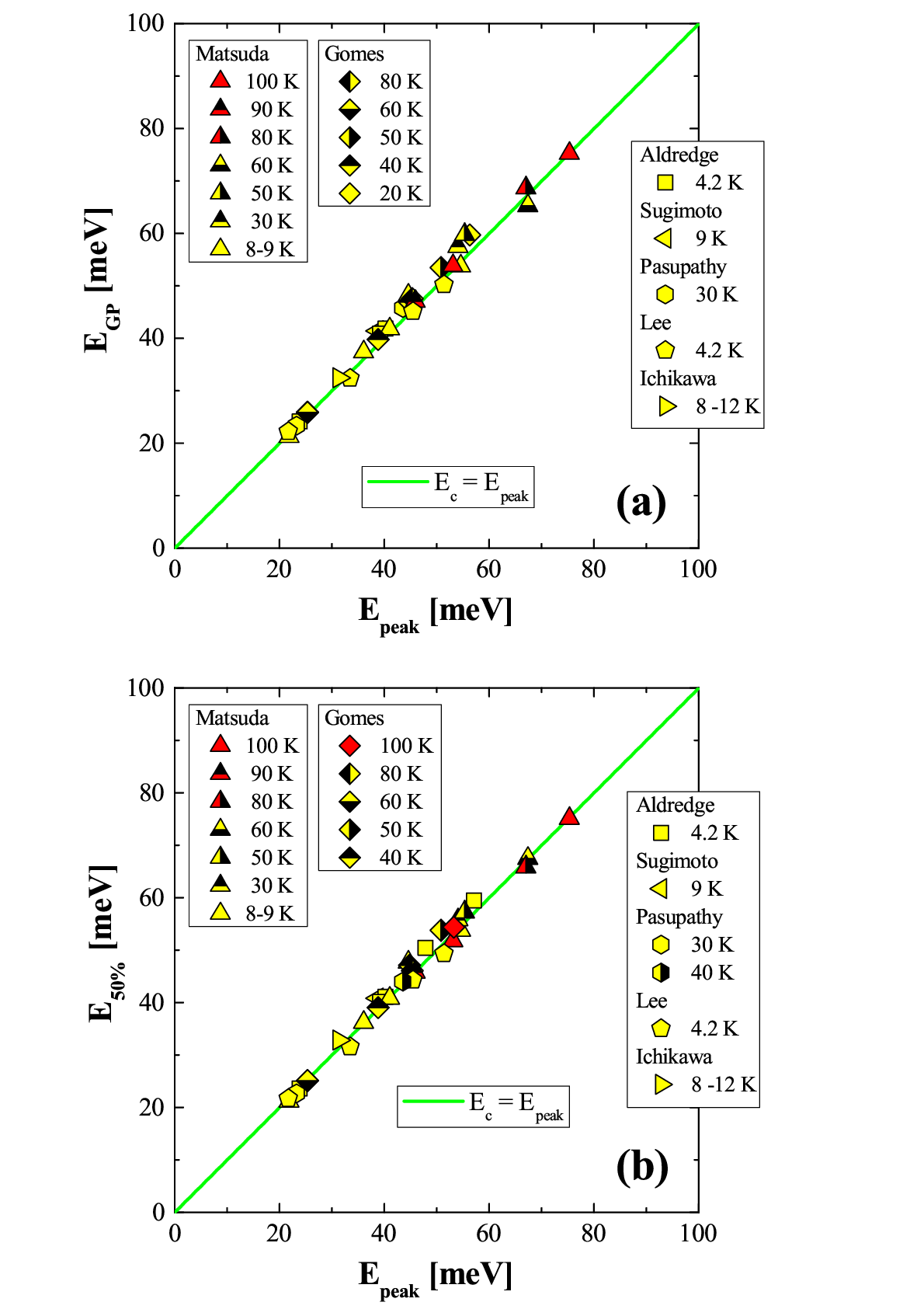}
\caption{\label{fig3} (a) The peak value ($E_{GP}$) of the fitted Gaussian distribution versus the peak value ($E_{peak}$) read directly from the gap distribution for the purely oxygen-doped Bi$_2$Sr$_2$CaCu$_2$O$_{8+\delta}$. (b) $E_{50\%}$ versus observed $E_{peak}$ for the purely oxygen-doped Bi$_2$Sr$_2$CaCu$_2$O$_{8+\delta}$. The yellow symbols are data measured for $T < T_c$, and the red symbols are data measured for $T_c < T < T_{100\%} = T_{lp}^*$. The plotted data are from Refs.\ \cite{gom07,mat03,lee06,sug06,all08,pas08,ich09}. }
\end{figure}

In Fig.\ \ref{fig3}(a), we plot the expectation value, i.e. the peak value ($E_{GP}$), of the fitted Gaussian distribution versus the peak value ($E_{peak}$) read directly from the gap distribution observed in Bi$_2$Sr$_2$CaCu$_2$O$_{8+\delta}$ by various groups \cite{gom07,mat03,lee06,sug06,all08,pas08,ich09} for  $T < T_{100\%} = T_{lp}^*$. It can be clearly seen that when we treat the gap distribution of whole INSEG map as a single Gaussian distribution, the $E_{GP}$ closely traces the $E_{peak}$. This validates, to the zeroth order, our choice of using a single Gaussian distribution to analyze INSEG maps.

In Fig.\ \ref{fig3}(b), we plot the $E_{50\%}$ versus $E_{peak}$ observed in Bi$_2$Sr$_2$CaCu$_2$O$_{8+\delta}$ for $T < T_{100\%} = T_{lp}^*$, respectively \cite{gom07,mat03,lee06,sug06,all08,pas08,ich09}. $E_{50\%}$ is almost the same as the $E_{peak}$ and, therefore, $E_{GP}$. Accordingly, the $E_{peak}$ and $E_{GP}$ are also corresponding to the upper pseudogap energy observed in the UEPD. Since the upper pseudogap temperature is observed by the dc resistivity, as shown in Fig.\ \ref{fig1}(a), which is a typical bulk probe, the present result of $E_{peak}$ = $E_{50\%}$ = $E_{GP}$ implies that the three energies $E_{peak}$, $E_{GP}$ and $E_{50\%}$ share the identical physical meaning of the expectation value measured by the experimental probes.

\begin{figure}[b]
\includegraphics [scale=0.4]{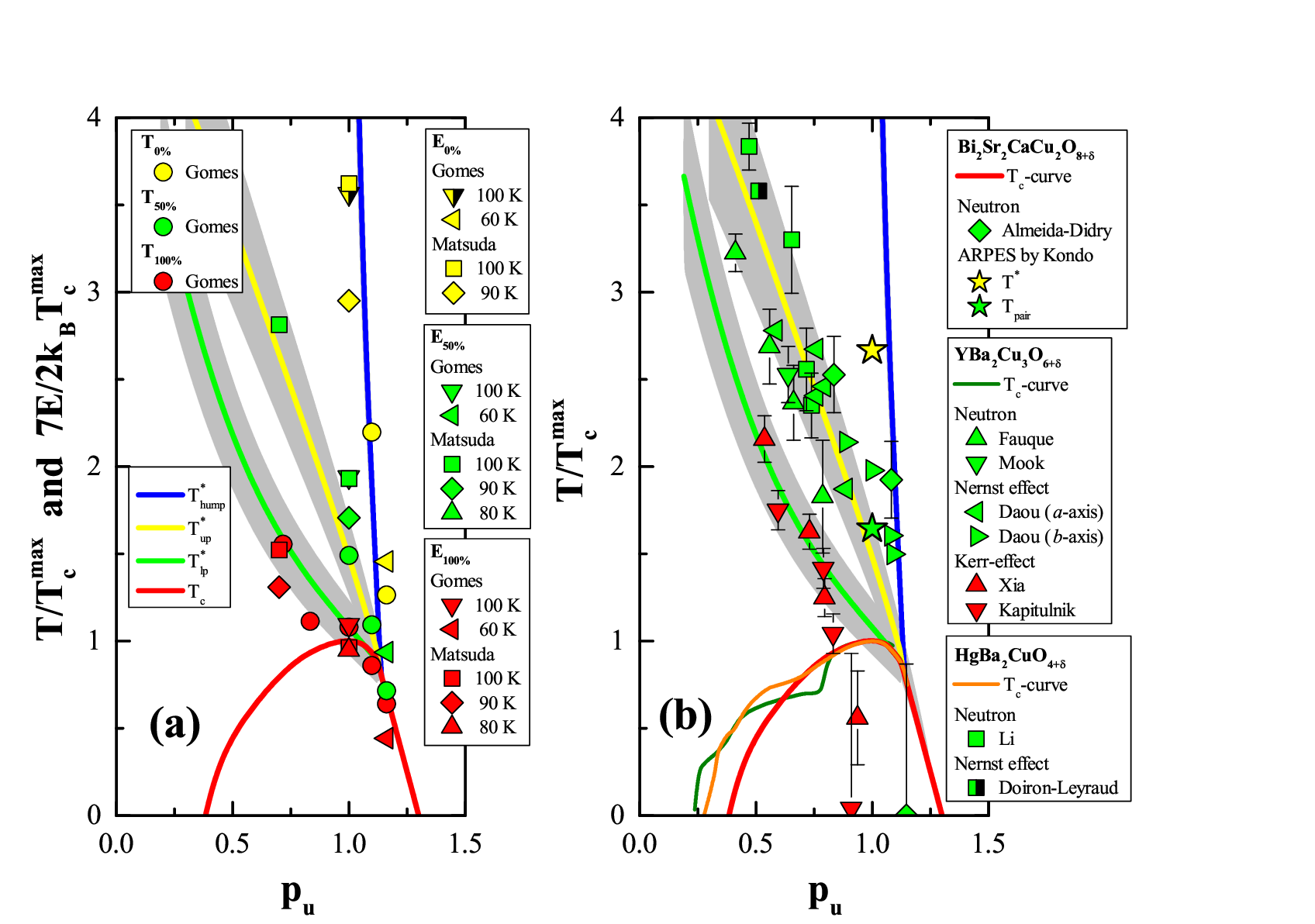}
\caption{\label{fig4} (a) UEPD with all characteristic temperatures and energies of the INSEG state in Bi$_2$Sr$_2$CaCu$_2$O$_{8+\delta}$. We use the same convention of  symbols as that in Figs.\ \ref{fig2}(b) and\ \ref{fig2}(c).  (b) UEPD with the recent data from Refs.\ \cite{fau06,moo08,xia08,kap09,dao10,li08,li11,ley12,did12,kon13} published after the publication of Ref.\ \cite{hon08} in 2008. The $T_c$-curve for YBa$_2$Cu$_3$O$_{6+\delta}$ is from Ref.\ \cite{hon07}. The $T_c$-curve for HgBa$_2$CuO$_{4+\delta}$ is obtained by analyzing the data from Ref.\ \cite{yam00} using $P_{pl}$-scale. }
\end{figure}

In the $P_{pl}$-scale the optimal doped-hole concentration $P_{pl}^{opt}$ depends on the individual HTCS material \cite{hon08}. However when using the reduced temperature $T/T_c^{max}$ (the reduced energy $2E/7k_BT_c^{max}$) and the reduced hole-concentration $p_u \equiv P_{pl}/P_{pl}^{opt}$ various HTCS can be easily compared with each other in spite of the variations in $P_{pl}^{opt}$ and $T_c^{max}$. Using $p_u$ and $T/T_c^{max}$ ($2E/7k_BT_c^{max}$), for HTCS with $T_c^{max} \sim$ 90 K, such as YBa$_2$Cu$_3$O$_{6+\delta}$, HgBa$_2$CuO$_{4+\delta}$ and Bi$_2$Sr$_2$CaCu$_2$O$_{8+\delta}$, various characteristic temperatures (energies) can be unified into a UEPD shown as the three lines with shaded area in Fig.\ \ref{fig4} \cite{hon08}. This UEPD was constructed based on the analysis of the experimental data measured by fifteen different macroscopic and microscopic experimental probes \cite{hon08}. Within these fifteen probes, five were surface-sensitive probes and ten were bulk probes. Therefore, UEPD represents a true intrinsic electronic phase diagram for HTCS with $T_c^{max} \sim$ 90 K.  We re-plot $T_{0\%}$, $T_{50\%}$ and $T_{100\%}$ in Fig.\ \ref{fig2}(b) and $E_{0\%}$, $E_{50\%}$ and $E_{100\%}$ in Fig.\ \ref{fig2}(c) into the Fig.\ \ref{fig4}(a). We confirm again that both the 50\% and 100\% coverage of CuO$_2$ planes are consistent with the intrinsic, universal upper and lower pseudogaps, respectively and the hump energy is corresponding to the onset ($E_{0\%}$) of the INSEG. In Fig.\ \ref{fig4}(b), we also include the most recent experimental results performed on YBa$_2$Cu$_3$O$_{6+\delta}$ \cite{fau06,moo08,xia08,kap09,dao10}, HgBa$_2$CuO$_{4+\delta}$ \cite{li08,li11,ley12}, and Bi$_2$Sr$_2$CaCu$_2$O$_{8+\delta}$ \cite{did12}. We can see that: (1) the polarized elastic neutron scattering experiments suggesting a novel translational-symmetry-preserving magnetic transition falls on $T_{up}^*$, (2) the Nernst effect measurements indicating a breaking of the 90$^\circ$-rotational ($C_{4v}$) symmetry occurs at $T_{up}^*$ and (3) the Kerr-effect measurements signaling a time-reversal symmetry breaking corresponds to $T_{lp}^*$. These experimental observations of very subtle changes of physical properties by different probes in different materials that universally fall on either the upper or the lower pseudogap reconfirm the two-pseudogap scenario reported in Ref.\ \cite{hon04}. This further validates the UEPD reported in Ref.\ \cite{hon08}. The novel connection between pseudogaps and nanogap distribution is a natural consequence of plotting the published experimental results by using the quantitatively correct and accurate $P_{pl}$-scale. Based on this inhomogeneous nanogap distribution picture and its quantitative connection to the bulk pseudogap, it is interesting to point out that these ``phase transitions'' are highly unusual that (1) and (2) appeared right at a 50\% coverage and (3) occurred when 100\% of the CuO$_2$ planes are covered by the nanogaps. How does a phase transition emerge from such an inhomogeneous background at the specific nanogap coverage require further studies.

In Fig.\ \ref{fig4}(b), we also include the most recent results observed in the optimally-doped Bi$_2$Sr$_2$CaCu$_2$O$_{8+\delta}$ of angle-resolved photoemission spectroscopy (ARPES) by using a new quantitative approach based on the temperature dependence of partial density of states at Fermi surface \cite{kon13}. The observed pseudogap temperature ($T^*$) and pair formation temperature ($T_{pair}$) lie on the hump and the upper pseudogap energy, respectively. Especially, their $T^*$ is the temperature when energy gap first appeared. This is exactly equal to our conclusion that hump energy is the onset of INSEG state, namely, the pseudogap they observed at $T^*$ is the onset of INSEG.

To understand why various probes can detect the pseudogap of either the 50\% or 100\% coverage of CuO$_2$ planes by INSEG, we point out that the relationship between $E^*$ and $T^*$, 2$E^*/k_B T^*$ = 7 $\pm$ 1, revealed in the UEPD plot, and that between $\Delta_g$ and $T_p$, 2$\Delta_g/k_BT_p$ = 7.9 $\pm$ 0.5 reported in Ref.\ \cite{gom07}, are surprisingly similar. The difference between the two relations is within the error band of the original construction of the UEPD. The UEPD was constructed by the data collected from many experimental techniques which probe an area with a length-scale that is much larger than the characteristic length-scale, $\sim$ 10$^{-9}$ $m$, of INSEG. This strongly suggests that the pseudogaps revealed in UEPD are the expectation value of the gap map sampled in the characteristic length-scale of the experimental probe. Indeed, it is intriguing to see that the gap map identified in the STM/S can be naturally related to the ``bulk'' pseudogap measured by various bulk probes: $E_{up}^*$ ($T_{up}^*$) and $E_{lp}^*$ ($T_{lp}^*$) are the gap energies (temperatures) when one half and the entire CuO$_2$ plane are covered by the INSEG, respectively. The former is detected by the downward deviation from the linear temperature dependence of in-plane dc resistivity on cooling at the high temperature \cite{ito93}, and Nernst effect \cite{dao10,ley12}. The latter are properties that are detected by various bulk and surface-sensitive probes \cite{hon08}. It is interesting to note that the 50\% coverage corresponds to 2D bond-percolation limit of a square lattice \cite{kes80}, which clearly indicates the in-plane conductivity change is due to percolation. Therefore, we conclude that the upper and lower pseudogaps detected by different experimental probes must also be related to, besides the characteristic length-scale, the energy-scale of the experimental probes. In this context, the pseudogaps in the UEPD are the spatially ``averaged'' response of the gap map measured by the individual experimental probe. Depending on the characteristic energy-scale and length-scale of the experimental probes: some of the probes are sensitive enough to pick up the incipient inhomogeneous nanoscale electronic state, some probes pick up the bond-percolation when completed and the others measure the bulk property when the CuO$_2$ planes are completely covered by INSEG.

There are two mutually exclusive scenarios regarding the connection between INSEG coverage and the pseudogaps: one is that the connection between 50\% (100\%) coverage to the upper (lower) pseudogap is only a surface manifestation of a intrinsically bulk pseudogap, therefore, the nanoscale inhomogeneity is just a surface state that is distinct from bulk. The other is that it is an intrinsic property of the CuO$_2$ plane that the INSEG is not confined to the surface but also exists throughout the bulk. To resolve these two conflicting view it is important to point out the fact that both pseudogap and the INSEG are pure 2D properties. Indeed recent studies showed that the in-plane charge ordering observed by surface and bulk probes are the same in the purely oxygen-doped Bi$_2$Sr$_2$CaCu$_2$O$_{8+\delta}$ \cite{net14}. Similar observations that electronic structure observed by STM exists in the bulk are also reported in the La-doped Bi$_2$Sr$_2$CuO$_{6+\delta}$ \cite{com14} and the Dy-doped Bi$_2$Sr$_2$CaCu$_2$O$_{8+\delta}$ \cite{net14}. These are clear indications that the electronic structure observed by STM on the surface also exists in the bulk. On the other hand we are not aware of any bulk transition that induces the surface coverage properties as we have observed here. It is very difficult to envision that the connection between the specific coverage of CuO$_2$ planes and the two pseudogaps over such wide doping range is only a surface property. In fact, similar characteristics and similar nanoscale inhomogeneity were observed in very different HTCS \cite{kat05,koh07} that we expect, if the gap map data sets are available in the literature, then the gap map would lead to the same coverage as we observed. Furthermore, the present quantitative connection between the INSEG observed in Bi$_2$Sr$_2$CaCu$_2$O$_{8+\delta}$ to the UEPD suggests that the INSEG state is a universal property for hole-doped HTCS with $T_c^{max} \sim$ 90 K. In light of the aforementioned observations and the further connection between the 100\% and 50\% coverage of CuO$_2$ planes by INSEG to the lower and the upper pseudogaps, respectively, we argue that the specific coverage of the CuO$_2$ planes by INSEG should be at least intrinsic to Bi$_2$Sr$_2$CaCu$_2$O$_{8+\delta}$ and, more likely, to be generic to HTCS with $T_c^{max} \sim$ 90 K.

High-$T_c$ superconductivity at around 100\% gap coverage for optimally and overdoped HTCS as can be clearly seen in Fig. 5 in Ref.\ \cite{gom07}. However superconductivity in the underdoped regime, as seen in Fig.\ \ref{fig2}, appears at an even lower temperature {\itshape\bfseries after} the CuO$_2$ planes are {\itshape\bfseries completely} covered by the INSEG. Therefore, combing all the above observations and for the entire doping range, we conclude that 100\% coverage of the CuO$_2$ planes by the INSEG is a {\itshape\bfseries necessary} condition for generating the high-$T_c$ superconductivity in cuprate superconductors. We emphasize that our conclusion is fundamentally different from other two-energy-scale scenarios where the pseudogap or charge order is competing against superconductivity. In contrast, we proposed that the 100\% coverage of the CuO$_2$ planes by the INSEG is a necessary condition for the high-$T_c$ superconductivity, and the high-$T_c$ superconductivity is ``realized''on a texture of a globally coupled INSEG of the lower pseudogap state.

\section{Conclusions}

The topographic coverage interpretation of the pseudogaps provides a microscopic inhomogeneous electronic picture for the origin of the pseudogap and superconductivity. Based on this picture, the pseudogaps, an observable due to the averaged response of the topographic coverage at 50\% or 100\% in the gap map detected by a specific experimental probe, loses its conventional meaning of a ``gap''.  It is in this context that the gap is ``pseudo'', and accordingly, all properties measured on HTCS should be addressed with the characteristic length-scale and energy-scale of the experimental probes, and the underlying INSEG state in mind \cite{sun06}. Indeed, the photon-energy-dependence of the ARPES  spectra were only recently observed \cite{has12}, when laser-based ARPES had achieved an unprecedented high-resolution, indicating that the probe energy should be as low as possible in addressing the low energy quasiparticle states. While our results link the specific INSEG coverage of CuO$_2$ planes to the two pseudogaps, the origin of and how the high-$T_c$ superconductivity emerges from such a robust INSEG state remain to be a challenging problem of the mechanism for HTCS.

\section*{Acknowledgement}

P.H.H. is supported by the State of Texas through the Texas Center for Superconductivity at University of Houston.


\begin{thebibliography}{00}


\bibitem{ito93} T. Ito, K. Takenaka, S. Uchida, Phys. Rev. Lett. 70 (1993) 3995.
\bibitem{din96} H. Ding et al., Nature 382 (1996) 51.
\bibitem{tim99} T. Timusku, B. Statt, Rep. Prog. Phys. 62 (1999) 61.
\bibitem{dam03} A. Damascelli, Z. Hussain, Z. -X. Shen, Rev. Mod. Phys. 75 (2003) 473.
\bibitem{fis07} \O. Fisher et al., Rev. Mod. Phys. 79 (2007) 353.
\bibitem{hon04} T. Honma et al., Phys. Rev. B 70 (2004) 214517.
\bibitem{hon06} T. Honma, P. H. Hor, Supercond. Sci. Technol. 19 (2006) 907.
\bibitem{add01} Please see papers of Refs.\cite{hon04,hon06,hon10} regarding the use of proper dimension of a hole-scale to reveal the true physical properties.
\bibitem{hon10} T. Honma, P. H. Hor, Physica C 470 (2010) S191.
\bibitem{hon08} T. Honma, P. H. Hor, Phys. Rev. B 77 (2008) 184520. 
\bibitem{add02} The hump energy is the energy of the so-called hump observed in the ARPES and SIS tunneling experiments. Also see Fig. 6(d) in Ref. \cite{hon08}.
\bibitem{pan01} S. H. Pan et al., Nature 413 (2001) 282. 
\bibitem{gom07} K. K. Gomes et al., Nature 447 (2007) 569.
\bibitem{boy07} M. C. Boyer et al., Nat. Phys. 3 (2007) 802. 
\bibitem{hon07} T. Honma, P. -H. Hor, Phys. Rev. B 75 (2007) 012508. 
\bibitem{yam00} A. Yamamoto, W. -Z. Hu, S. Tajima, Phys. Rev. B 63 (2000) 024504.
\bibitem{mat03} A. Matsuda, F. Takenori, T. Watanabe, Physica C 388-389 (2003) 207. 
\bibitem{lee06} J. Lee et al., Nature 442 (2006) 546. 
\bibitem{sug06} A. Sugimoto et al., Phys. Rev. B 74 (2006) 094503. 
\bibitem{all08} J. W. Alldredge et al., Nat. Phys. 4 (2008) 319.
\bibitem{pas08} A. N. Pasupathy et al., Science 320 (2008) 196. 
\bibitem{ich09} H. Ichikawa et al., Physica C 469 (2009) 1013. Here, only Bi$_2$Sr$_2$CaCu$_2$O$_{8+\delta}$ data is used for the present analysis.
\bibitem{fau06} B. Fauqu\'{e} et al., Phys. Rev. Lett. 96 (2006) 197001.
\bibitem{moo08} H. A. Mook et al., Phys. Rev. B 78 (2008) R020506.
\bibitem{xia08} J. Xia et al., Phys. Rev. Lett. 100 (2008) 127002.
\bibitem{kap09} A. Kapitulnik et al., New J. Phys. 11 (2009) 055060.
\bibitem{dao10} R. Daou et al., Nature (London) 463 (2010) 519.
\bibitem{li08} Y. Li et al., Nature (London) 455 (2008) 372.
\bibitem{li11} Y. Li et al., Phys. Rev. B 84 (2011) 224508. 
\bibitem{ley12} N. Doiron-Leyraud et al., Phys. Rev. X 3 (2013) 021019.
\bibitem{did12} S. De Almeida-Didry et al., Phys. Rev. B 86 (2012) R020504.
\bibitem{kon13} T. Kondo et al., Phys. Rev. Lett. 111 (2013) 157003.
\bibitem{kes80} H. Kesten, Commun. Math. Phys. 74 (1980) 41. 
\bibitem{net14}E. H. da Silva Neto et al., Science 343 (2014) 393.
\bibitem{com14} R. Comin et al., Science 343 (2014) 390.

\bibitem{kat05} T. Kato, S. Okitsu, H. Sakata, Phys. Rev. B 72 (2005) 144518.
\bibitem{koh07} Y. Kohsaka et al., Science 315 (2007) 1380.

\bibitem{sun06} X. F. Sun et al., Phys. Rev. Lett. 96 (2006) 017008.
\bibitem{has12} M. Hashimoto et al., Phys. Rev. B 86 (2012) 094504.

\end{thebibliography}
\end{document}